\documentclass[a4paper,twosides]{article}
\usepackage{multicol,multirow,graphicx,fancyhdr,epstopdf}
\usepackage{amsmath,amsfonts,amssymb,bm,upgreek,mathrsfs,ccmap,mathcomp}
\usepackage[pagewise,switch,columnwise]{lineno}
\usepackage[compress,nospace]{cite}
\usepackage[dvipsnames]{xcolor}
\usepackage{CPL-2022}

\newcommand{\cplyear}{x} \newcommand{\cplvol}{xx}
\newcommand{\cplno}{x} \newcommand{\cplpagenumber}{xxxxxx}
 \newcommand{\cplpage}{\cplpagenumber-\thepage}

\newcommand{\mev}{\mathrm{MeV}} 
\newcommand{\kev}{\mathrm{keV}} 
\newcommand{\gev}{\mathrm{GeV}} 
\newcommand{\ev}{\mathrm{eV}}

\begin{document}

\vspace* {-4mm} 
\begin{center}
\large\bf{\boldmath{A short review of the vector charmonium-like state $\psi(4230)$}}
\footnotetext{\hspace*{-5.4mm}$^{*}$Corresponding authors. Email: qianwang@m.scnu.edu.cn; zhaoq@ihep.ac.cn

\noindent\copyright\,{\cplyear}
\href{http://www.cps-net.org.cn}{Chinese Physical Society} and
\href{http://www.iop.org}{IOP Publishing Ltd}}
\\[5mm]
\normalsize \rm{}Qian Wang$^{1,3,5,6*}$, and Qiang Zhao$^{2,4,7*}$
\\[3mm]\small\sl

$^{1}$ State Key Laboratory of Nuclear Physics and Technology, Institute of Quantum Matter, \\ South China Normal University, Guangzhou 510006, China

$^{2}$ Institute of High Energy Physics, Chinese Academy of Sciences, Beijing 100049, China

$^{3}$ Guangdong Basic Research Center of Excellence for Structure and Fundamental Interactions of Matter, \\ Guangdong Provincial Key Laboratory of Nuclear Science, Guangzhou 510006, China

$^{4}$ University of Chinese Academy of Sciences, Beijing 100049, China

$^{5}$ Research Center for Nuclear Physics (RCNP), Osaka University, Ibaraki 567-0047, Japan

$^{6}$ Southern Center for Nuclear-Science Theory (SCNT), Institute of Modern Physics, Chinese Academy of Sciences, Huizhou 516000, Guangdong Province, China

$^{7}$ Center for High Energy Physics, Henan Academy of Sciences, Zhengzhou 450046, China

\leavevmode\\[1mm]\normalsize\rm{}(Received xxx; accepted manuscript online xxx)
\end{center}
\vskip 1.5mm

\small{\narrower We present a concise review of the vector charmonium state $\psi(4230)$, which was originally labelled as $Y(4260)$ in the literature. As one of the earliest candidates for a QCD exotic states, its interpretation has initiated various ideas about possible manifestations of non-perturbative mechanisms in the charmonium mass regime. In this short article we briefly review the experimental status of $\psi(4230)$ and discuss possible theoretical interpretations. We will focus on four broadly investigated scenarios, i.e.  tetraquark, hybrid, hadro-quarkonium, and hadronic molecule, and highlight the key issues based on these approaches. Crucial experimental observables, e.g. mass position, lineshapes, di-lepton decay width $\Gamma_{ee}$, production rates in $B$ meson decays, dominant hadronic decay patterns, and the potential $1^{-+}$ and $0^{--}$ exotic partners, are assessed, which can provide crucial structure information for understanding this mysterious state. 

\par}\vskip 3mm
{\bf Keywords:} Exotic hadrons, vector charmonium-like state, Heavy Quark Spin Symmetry partner\\
\noindent{\narrower{DOI: \href{http://dx.doi.org/10.1088/0256-307X/\cplvol/\cplno/\cplpagenumber}{10.1088/0256-307X/\cplvol/\cplno/\cplpagenumber}}

\par}\vskip 5mm
\section{Introduction}
Although Quantum Chromo-dynamics (QCD) has proven to be the correct theory for strong interaction, 
we still have not yet fully understood the mechanism, namely, 
how quarks and gluons are confined to form color-singlet hadrons.  For the simplest color-singlet hadrons, i.e. mesons composed of quark-antiquark ($q\bar{q}$) and baryons of three quarks ($qqq$), the conventional quark model provides an economic and successful prescription for understanding their classifications and mass relations with the ``so-called" constituent quarks as their ingredients. 

In the light flavor sector the conventional quark model becomes qualitative. This is because the masses of the light constituent quarks, which are about $300\sim 400$ MeV for the $u/d$ flavors and about $450\sim 550$ MeV for the strange one, are comparable with the excitation energy scale for the light hadrons which are about $500\sim 600$ MeV. For instance, the mass difference between the nucleon and the first orbital excitation non-strange baryon $S_{11}(1535)$ with $J^{P}=1/2^-$ is about 600 MeV~\cite{ParticleDataGroup:2024cfk}. This means that it is hard to justify the non-relativistic feature of such a system, and the mass spectra obtained by solving the Schr\"odinger equation will bare large intrinsic uncertainties. In contrast, the situation improves for systems containing heavy quarks. In particular, for the heavy flavor hadrons such as the charmonia and bottomonia, with the constituent quark masses much larger than the typical energy level of the excited states, more quantitative description of the mass spectra can be obtained. As shown by the Cornell model~\cite{Eichten:1974af,Eichten:1978tg,Eichten:1979ms}, the low-lying charmonium and bottomonium spectra below the $S$-wave open-flavor threshold can be well described by the quark potentials which include the linear confinement and the Coulomb part of the one-gluon exchange potential. With the spin-dependent interactions the quark model succeeded in describing the low-lying charmonium and bottomonium spectra with rather high precision~\cite{Godfrey:1985xj}. 

Apart from the simple quark-model scenarios for conventional hadrons, the non-Abelian property of QCD does not prohibit the formation of color-singlets from more complicated constituent structures. Such states are beyond the $(q\bar{q})$ color singlet for meson, or $(qqq)$ for baryon in the conventional quark model picture, and often labeled as ``QCD exotics", such as tetraquark, pentaquark, glueball, hybrid, and hadronic molecule, etc. Experimental search and theoretical study of these exotic hadrons will provide crucial information for understanding the non-perturbative property of QCD.

The first strong evidence for the failure of the simple quark-model picture for the heavy quarkonium spectra was due to the observation of the $X(3872)$, aka $\chi_{c1}(3872)$, in 2003~\cite{Belle:2003nnu}. As the candidate of the $\chi_{c1}(2P)$ state, its mass is far below the expectation of any known quark model predictions~\cite{Eichten:1974af,Eichten:1978tg,Eichten:1979ms,Godfrey:1985xj,Lakhina:2006vg}. Since 2003, with more and more data from experiment, tens of new resonance-like structures have been observed. Some of these apparently deviate from the conventional quark model expectations, and thus, make them ideal candidates for QCD exotics beyond the conventional quark model. 
The properties of some of these states have been  intensively studied from various perspectives. For the status of experimental and theoretical progresses, one can refer to  several recent review articles~\cite{Swanson:2006st,Chen:2016qju,Hosaka:2016pey,Richard:2016eis,Lebed:2016hpi,Esposito:2016noz,Guo:2017jvc,Ali:2017jda,Olsen:2017bmm,Liu:2019zoy,Brambilla:2019esw,Yamaguchi:2019vea,Guo:2019twa,Yang:2020atz,Yuan:2021wpg,Chen:2022asf,Mai:2022eur,Meng:2022ozq,Liu:2024uxn,Huang:2023jec,Montana:2023sft,Chen:2024eaq,Mai:2025wjb,Esposito:2025hlp,Lu:2025syk,Albuquerque:2018jkn,Nielsen:2009uh,Wang:2025sic}.

So far, most heavy quarkonium-like states were observed in either $e^+e^-$ collider or hadron collider experiments, for instance, by Belle, BaBar, BESIII, LHCb, and CLEO-c Collaborations. Those exotic candidates can be produced from either prompt productions or decays of higher states. For the vector charmonium- or bottomonium-like states,  the advantage is that they can be directly produced in $e^+e^-$ annihilation. Thus, the $e^+e^-$ collider also provides a direct examination of the theoretical predictions for the vector charmonium and/or bottomonium spectra. 

The experimental measurement of the vector charmonium spectra confirms the success of the conventional quark model for the low-lying states. Below 4.2 GeV the observed vector spectrum can be well described by the potential quark model calculations~\cite{Eichten:1978tg,Eichten:1979ms,Godfrey:1985xj,Lakhina:2006vg}. Namely, the low-lying vectors, $J/\psi \ (\psi(1S))$, $\psi(3686)(2S)$, $\psi(3770)(1D)$, $\psi(4040)(3S)$, $\psi(4160)(2D)$, can be assigned to the eigen states calculated by the potential quark model.  In 2005, BaBar collaboration first reported a vector charmonium state $Y(4260)$ in $e^+e^-$ annihilation via the initial-state-radiation (ISR) process $e^+e^-\to \gamma_\mathrm{\mathrm{ISR}}\pi^+\pi^-J/\psi$~\cite{BaBar:2005hhc}, which cannot be accommodated by the potential quark model for $c\bar{c}$. Since then, a number of higher vector charmonium states were reported in experiment with the increasing data accumulated by Belle, BaBar and BESIII Collaborations, e.g. $Y(4008)$,  $Y(4360)$, $Y(4660)$, etc~\cite{ParticleDataGroup:2024cfk}. 

In 2013, with a large data sample from energy scan, BESIII Collaboration found the tetraquark candidate $Z_c(3900)$ in  $e^+e^-\to Y(4260)\to J/\psi\pi^+\pi^-$~\cite{BESIII:2013ris}, which suggests that the production of $Z_c(3900)$ is strongly correlated with the unusual property of $Y(4260)$, making $Y(4260)$ unique in various charmonium-like states. The data also show that $Y(4260)$ has a nontrivial cross section lineshape, with  
its mass position shifted to be around $(4.217 \pm 2.0)~\mathrm{GeV}$~\cite{Cleven:2013mka}, and
the nearby $D_1\bar{D}$ threshold accounts for the largely asymmetric lineshape of the $Y(4260)$. 
This contradicts to the originally measured larger mass extracted from one symmetric Breit-Wigner fitting.  Alternatively, one can also use two coherent Breit-Wigner (B-W) states to describe this highly asymmetric lineshape. In the $h_c\pi^+\pi^-$ channel the analysis indicates the existence of a narrow state around $4.22~\mathrm{GeV}$~\cite{Yuan:2013uta}. 
Improved precision measurement in $h_c\pi^+\pi^-$~\cite{BESIII:2016adj} further confirms the existence of $Y(4220)$. This narrow structure is also seen in $e^+e^-\to\omega \chi_{c0}$ by BESIII~\cite{BESIII:2014rja} Collaboration, which seems to be consistent with the lower state at $(4222.0\pm 3.1)~\mathrm{MeV}$ fitted from the fine structure of $Y(4260)$ in $e^+e^-\to J/\psi\pi^+\pi^-$ with two B-W resonances~\cite{BESIII:2016bnd}. In the fitting with two coherent resonances in Refs.~\cite{BESIII:2016bnd,BESIII:2016adj} it shows that the lower one has a relatively narrow width with a mass around $4.22~\mathrm{GeV}$ and the higher one is broad with a mass around $4.29\sim 4.32~\mathrm{GeV}$. Note that the higher one observed in $h_c\pi^+\pi^-$ is known as $Y(4360)$. 
As shown by the high-statistics data mentioned above, the peak of the $Y(4260)$ has shifted to be around $4.23~\mathrm{GeV}$. 
In the new naming scheme of charmonium~\cite{ParticleDataGroup:2014cgo} all the vector charmonium-like states are named as $\psi$ states. Therefore, $Y(4260)$ is renamed as $\psi(4230)$, and we use $\psi(4230)$ to denote $Y(4260)$ (or $Y(4230)$) in the rest of this article.

Due to these interesting observations and unusual behaviors of $\psi(4230)$ we are motivated to review its property and try to gain some insights into the underlying dynamics.  
As follows, we first give a brief review of the experimental history concerning the observations of $\psi(4230)$ in Sec.~\ref{sec-2}. In Sec.~\ref{sec-3} we discuss the theoretical interpretations of $\psi(4230)$ based on different scenarios. Crucial issues concerning the properties of $\psi(4230)$ will be stressed. We emphasize that to better understand the nature of $\psi(4230)$, it should be necessary to accommodate all these issues in a self-contained framework. A brief summary will be given in the last Section.

\section{Experimental evidences }\label{sec-2}

As mentioned earlier, $\psi(4230)$ was first reported by BaBar Collaboration~\cite{BaBar:2005hhc} in the ISR process $e^+e^-\to \gamma_\mathrm{ISR}\pi^+\pi^-J/\psi$, and then quickly confirmed by CLEO~\cite{CLEO:2006ike} and Belle~\cite{Belle:2007dxy} Collaborations. During the past two decades there have been continuous experimental efforts on studying this state in different processes. In particular, $e^+e^-$ annihilation provides a direct probe and the BESIII experiment has accumulated a large data sample via energy scan. Apart from the exclusive channel of $e^+e^-\to J/\psi \pi^+\pi^-$ which is the discovery channel of $Z_c(3900)$~\cite{BESIII:2013ris,BESIII:2022qal}, signals of $\psi(4230)$ were observed in $e^+e^-\to J/\psi\pi^0\pi^0$~\cite{BESIII:2020oph}, $J/\psi K^+K^-$~\cite{BESIII:2022joj}, $J/\psi\eta$~\cite{BESIII:2020bgb}, $\chi_{c1}(1P)\omega$~\cite{BESIII:2019gjc}, $D^0D^{*-}\pi^+$~\cite{BESIII:2018iea}, $\chi_{c1}(3872)\gamma$~\cite{BESIII:2019qvy}, $h_c(1P)\pi^+\pi^-$~\cite{BESIII:2016adj}, $\psi(2S)\pi^+\pi^-$~\cite{BESIII:2021njb}, and $\mu^+\mu^-$~\cite{Ablikim:2020jrn}. Recently, BESIII Collaboration also reported evidence for $\psi(4230)$ in $e^+e^-\to \Omega\bar{\Omega}$~\cite{BESIII:2025azv}, which may initiate further interesting questions concerning its internal structure. 

Since the discovery of $\psi(4230)$ there have been a lot of theoretical efforts on understanding its nature, for which we will discuss several broadly investigated scenarios in the next Section. In this Section we summarize the key experimental observables which should be crucial for disentangling the nature of $\psi(4230)$. 

\begin{enumerate}
    \item The cross section lineshape turns out to be one of the most important observables for understanding $\psi(4230)$. The lineshape can be measured directly in $e^+e^-$ annihilation and it manifests the interference between the resonance structure and either a background process or an additional resonance. For an ideal situation where the resonance is narrow and isolated from other nearby states or thresholds, one can directly extract the resonance parameters, e.g. mass and width. However, the situation becomes complicated if a state is located near an open threshold, deeply inside the energy complex plane, or has large overlaps with the nearby states of the same quantum numbers. In such cases, the cross section lineshape will be distorted, and an elaborate treatment is needed to extract the resonance parameters~\cite{Guo:2017jvc}.
    
    As mentioned earlier, the resonant lineshapes around $4.2-4.3~\mathrm{GeV}$ in more than ten channels have been measured by the BESIII Collaboration with the high-luminosity energy scan~\cite{BESIII:2020oph,BESIII:2020oph,BESIII:2022qal,BESIII:2022joj,BESIII:2020bgb,BESIII:2019gjc,BESIII:2018iea,BESIII:2019qvy,BESIII:2016adj,BESIII:2021njb,Ablikim:2020jrn}. As shown by the data, the lineshapes are significantly channel-dependent, making the extracted B-W masses different from each other. The extracted  masses in the above ten channels are collected in Fig.~\ref{fig:mass} to compare with the average value of the PDG~\cite{ParticleDataGroup:2024cfk}. The data show that a B-W fitting of $\psi(4230)$ does not work here taking into account that its mass is in the vicinity of the nearby $D_1\bar{D}$ and $D_0\bar{D}^*$ thresholds~\footnote{$D_0$ by itself has a not-well-located two-pole structure.}. The strong $S$-wave interaction can distort the lineshape as discussed in e.g. Ref.~\cite{Guo:2017jvc}.

    \begin{figure}
    \centering
\includegraphics[width=0.5\linewidth]{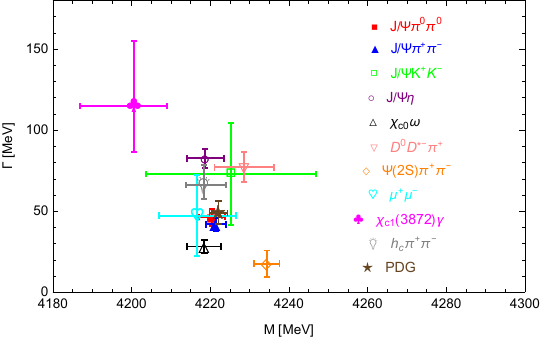}
    \caption{The Breit-Wigner masses and widths extracted in various channels are collected in this figure. The red box, blue triangle, green hollow box, purple hollow circle, black hollow triangle, pink hollow inverted triangle, orange hollow rhombus, light blue hollow heart, Megenta plum blossom, and gray light bulb are for those extracted in the $J/\psi\pi^0\pi^0$~\cite{BESIII:2020oph}, $J/\psi\pi^+\pi^-$~\cite{BESIII:2022qal}, $J/\psi K^+K^-$~\cite{BESIII:2022joj}, $J/\psi\eta$~\cite{BESIII:2020bgb}, $\chi_{c0}(1P)\omega$~\cite{BESIII:2019gjc}, $D^0D^{*-}\pi^+$~\cite{BESIII:2018iea}, $\psi(2S)\pi^+\pi^-$~\cite{BESIII:2021njb}, $\mu^+\mu^-$~\cite{Ablikim:2020jrn}, $\chi_{c1}(3872)\gamma$~\cite{BESIII:2019qvy}, $h_c(1P)\pi^+\pi^-$~\cite{BESIII:2016adj} channels, respectively. The brown star is the average value from PDG~\cite{ParticleDataGroup:2024cfk}. }   
    \label{fig:mass}
\end{figure}

    \item One of the mysterious features of $\psi(4230)$ is its vague signal in two-body open-charm (i.e. $D\bar{D}$, $D\bar{D}^*+c.c.$, $D^*\bar{D}^*$ channels) decay channels and in the inclusive hadronic cross sections of $e^+e^-\to \mathrm{hadrons}$~\cite{ParticleDataGroup:2024cfk}. Before the availability of the high-luminosity energy scan, the cross section lineshape in the vicinity of $\psi(4230)$ did not manifest a resonance structure. It could be either negligibly small or have a peculiar way of interference with other processes~\cite{Xue:2017xpu}. With the high-statistics data from BESIII Collaboration, the $R$-value lineshape at the mass of $\psi(4230)$ seems to have a dip structure instead of a peak~\footnote{The $R$-value refers to the cross section ratios of $e^+e^-\to \mathrm{hadrons}$ over $e^+e^-\to \mu^+\mu^-$ as a function of the center-of-mass energy. Also, see Ref.~\cite{ParticleDataGroup:2024cfk} for the $R$-value review.}. 
    
    The upper limit of the di-lepton width $\Gamma_{ee}(\psi(4230))<580~\ev$ at $90\%$ C.L. is extracted by fitting the $R$ values. within the energy region $[3.7,5.0]~\mathrm{GeV}$ ~\cite{Mo:2006ss} measured by BES collaboration. 
    This indicates a large decay width of $\psi(4230)\to J/\psi\pi^+\pi^-$ if its decays into open charm channels are small~\cite{Mo:2006ss}. Meanwhile, an overall analysis of all the experimental data indicates that the leptonic decay width of $\psi(4230)$ ranges from $\mathcal{O}(10^2)~\ev$ to $\mathcal{O}(1)~\kev$~\cite{Cao:2020vab}. It means that an acceptable interpretation should explain which channel contributes to the width of $\psi(4230)$ and what is its relation with the observed decay channel of $\psi(4230)\to J/\psi\pi^+\pi^-$.
    
    \item The BaBar Collaboration~\cite{BaBar:2005hhc} also reported a signal of $\psi(4230)$ in the exclusive decay of $B^\pm\to K^\pm J/\psi \pi^+\pi^-$  with the product of the branching fractions $\mathcal{B}(B^-\to K^-\psi(4230))\mathcal{B}(\psi(4230)\to J/\psi\pi^+\pi^-)=(2.0\pm 0.7\pm 0.2)\times 10^{-5}$. Belle Collaboration~\cite{Belle:2019pfg} only finds excesses of events above the background level with a significance of 2.1 and 0.9 standard deviations, respectively, for the charged and neutral $B\to K\psi(4230)$ decays. The upper limits of the product of the branching fractions are $\mathcal{B}\left(B^{+} \to \psi(4230) K^{+}\right) \times \mathcal{B}(\psi(4230) \to J/\psi\pi^+\pi^-)<1.4\times 10^{-5}$ and $\mathcal{B}\left(B^{0} \to \psi(4230) K^{0}\right) \times \mathcal{B}(\psi(4230) \to J/\psi\pi^+\pi^-)<1.7\times 10^{-5}$ at $90\%$ confidence level~\cite{Belle:2019pfg}. The signal of $\psi(4230)$ has been confirmed in both $J/\psi\pi^+\pi^-$ and $J/\psi\eta$ channels in $e^+e^-$ annihilation with comparable cross sections. In contrast, the LHCb analysis of $B^+\to K^+\psi(4230)$ with $\psi(4230)\to J/\psi\eta$ does not show the signal of $\psi(4230)$~\cite{LHCb:2022oqs}. In addition, based on its Run II data, D0 collaboration observes the sequential decay of the $b$-flavored hadrons $H_b\to \psi(4230)+\mathrm{anything}$, with $\psi(4230)\to Z_c(3900)^\pm\pi^\mp\to J/\psi\pi^\pm\pi^\mp$ ~\cite{D0:2019zpb}.
A continued search of $\psi(4230)$ should be useful for understanding its production mechanism. 
    
    \item Decay modes can provide important information about the internal structure of a given particle. As mentioned earlier, the open charm and hidden charm decays of $\psi(4230)$ behave differently. 
    The BaBar collaboration set the upper limits   
        ${\mathcal{B}(\psi(4230) \rightarrow D^* \bar{D})}/{\mathcal{B}(\psi(4230) \rightarrow J / \psi \pi^{+} \pi^{-})}<34$ and ${\mathcal{B}(\psi(4230) \rightarrow D^* \bar{D}^*)}/ \\ {\mathcal{B}(\psi(4230) \rightarrow J / \psi \pi^{+} \pi^{-})}<40$,
    at the $95\%$ confidence level~\cite{BaBar:2009elc}. With the high-statistics data samples, BESIII Collaboration provides the cross section ratios at the mass of $\psi(4230)$ for $e^+e^-\to D^0D^{*-}\pi^+$~\cite{BESIII:2018iea} to $e^+e^-\to J/\psi\pi^+\pi^-$~\cite{BESIII:2018iea} and $e^+e^-\to J/\psi K^+K^-$~\cite{BESIII:2022joj}, respectively, 
    \begin{eqnarray}
\frac{\sigma\left(e^+e^-\to D^0D^{*-}\pi^+\right)}{\sigma\left(e^+e^-\to J/\psi\pi^+\pi^-\right)}\simeq 3 \ ,
    \end{eqnarray}
    and
    \begin{eqnarray}
\frac{\sigma\left(e^+e^-\to D^0D^{*-}\pi^+\right)}{\sigma\left(e^+e^-\to J/\psi K^+K^-\right)}\simeq 50 \ .       
    \end{eqnarray}
    Similar quantities can be extracted for those exclusive reaction channels,  which will further constrain the properties of $\psi(4230)$.
 
    \item Given the unusual properties of $\psi(4230)$, questions on their dynamical consequences will be raised, whatever the possible mechanisms could be. For instance, in the hadronic molecule scenario the existence of the $(D_1\bar{D}-D\bar{D}_1)/\sqrt{2}$ molecule with $J^{PC}=1^{--}$ would imply the associated exotic state $(D_1\bar{D}+D\bar{D}_1)/\sqrt{2}$ with $J^{PC}=1^{-+}$~\cite{Wang:2014wga,Dong:2019ofp,Zhang:2025gmm}, whose existence depends on the model assumptions of the dynamics. Similarly, in the picture of hybrid state with quark-gluon constituents, the ground vector charmonium hybrid with $J^{PC}=1^{--}$ implies the existence of its spin partners with $J^{PC}=0^{-+}, \ 1^{-+}, \ 2^{-+}$~\cite{Isgur:1984bm,Dudek:2009qf}. Experimental search for such partner states and/or consequential phenomena are necessary for examining the validity of any theoretical solution. We will come back to more detailed discussions later.

\end{enumerate}

\section{Theoretical developments}\label{sec-3}

In the early stage, people tried to assign $\psi(4230)$ as a conventional charmonium state predicted by early studies~\cite{Ding:1995he,Ding:1993uy} or accommodate it in updated calculations~\cite{Li:2009zu,Anwar:2016mxo}. 
A relativistic quark model calculation~\cite{Llanes-Estrada:2005qvr} assigned it as the $\psi(4S)$ charmonium. In Ref.~\cite{Swanson:2005tq} it was found that the $\psi(3D)$  vector charmonium (with mass of $4460~\mev$) is much heavier than $\psi(4230)$. The unquenched quark model calculations also do not support its being a conventional charmonium~\cite{Liu:2014spa,Kanwal:2022ani}. 
Interestingly, Lattice QCD (LQCD) simulations also disfavor $\psi(4230)$ being  a conventional charmonium~\cite{Bali:2011rd}. 

Various possibilities were proposed in the literature. As usual, a state that cannot be accommodated by the conventional spectrum could be an ideal candidate for ``exotic" states. 
Maiani {\it et al.}~\cite{Maiani:2005pe} assigned it as a $c\bar{c}s\bar{s}$ tetraquark state with $D_s\bar{D}_s$ as its predominant decay channel. Meanwhile, one of the most attractive possibilities is the hybrid solution, i.e.  $c\bar{c}g$ hybrid, which was proposed by Refs.~\cite{Zhu:2005hp,Close:2005iz,Kou:2005gt}. Such a scenario was also studied by LQCD~\cite{Chen:2016ejo,Berwein:2015vca}, 
where the charmonium hybrid state with $J^{PC}=1^{--}$ is located around $4.2\sim 4.4$ GeV. 

The consequence of multiquark exotics and hybrid states is that one state actually implies a whole multiquark spectrum. To avoid this, alternative solutions were also investigated. In Ref.~\cite{Ding:2008gr} it was proposed that the $D_1\bar{D}+c.c.$ and $D_0\bar{D}^*+c.c.$ coupled-channel interactions can form a hadronic molecular state around the mass of 4.26 GeV. Note that
both $D_1(2430)$ and $D_0^*(2300)$ are too broad to produce a narrow structure. Besides that, both of them have two pole structures~\cite{Guo:2006fu,Guo:2006rp,Albaladejo:2016lbb}, necessitating the proper treatment of all relevant cuts~\cite{Filin:2010se,Guo:2011dd}. It raises questions on the binding mechanism. Later, the observation of the charged charmonium state $Z_c(3900)$ at BESIII Collaboration~\cite{BESIII:2013ris} provided crucial information for $\psi(4230)$. It was proposed in Ref.~\cite{Wang:2013cya} that $\psi(4230)$ can be produced by the interaction between the narrow $D_1(2420)$ and $\bar{D}$. 
The narrow $D_1(2420)$ and broad $D_1(2430)$ are the mixing states of the first orbital excitation states $|^1P_1\rangle$ and $|^3P_1\rangle$ in the quark model~\cite{Close:2005se}. Meanwhile, these two states can be written in terms of the spin-parity of the light quark degrees of freedom in language of heavy quark spin symmetry (HQSS), i.e.
\begin{equation}
\left (\begin{array}{c}
|D_1^\prime (2430)\rangle \nonumber\\
|D_1(2420)\rangle\\
\end{array}
\right )=\left (\begin{array}{c}
|1^+,j_l^p=\frac{1}{2}^+\rangle \nonumber\\
|1^+,j_l^p=\frac{3}{2}^+\rangle\\
\end{array}
\right )=\left (\begin{array}{cc}
\cos{\theta}& -\sin{\theta} \nonumber\\
\sin{\theta}& \cos{\theta}\\
\end{array}
\right )\left (\begin{array}{c}
|^1P_1\rangle \nonumber\\
|^3P_1\rangle\\
\end{array}
\right ),
\end{equation}
where in the heavy quark limit the mixing angle $\theta$ takes the ideal mixing angle $\theta_0 = -\arctan(\sqrt{2})=-54.7^\circ$. As a result, the physical state $D_1(2420)$, which is assigned to be the $|1^+,j_l^p=\frac{3}{2}^+\rangle$ state, will decay into $D^*\pi$ via a $D$-wave, and thus, becomes narrow. In contrast, $D_1(2430)$ can decay into $D^*\pi$ via an $S$-wave, which makes its width very broad. However, one also recognizes that the charm quark mass is not heavy enough to meet the heavy quark mass limit of $m_Q\to \infty$. Therefore, some HQSS breaking effects are anticipated~\cite{Xue:2017xpu}. 
Extensive investigations in line with this scenario can be found in Refs.~\cite{Wang:2013kra,Cleven:2013mka,Wu:2013onz,Qin:2016spb,Chen:2016byt,Wang:2016wwe,Cleven:2016qbn,Lu:2017yhl,Xue:2017xpu,vonDetten:2024eie,Duan:2024zuo}, as well as its two-pole structures\cite{Guo:2006rp}.

In Ref.~\cite{Chen:2010nv} it was proposed that the $\psi(4230)$ structure in $e^+e^-\to J/\psi\pi^+\pi^-$ was created by the nearby resonance interferences and not a genuine resonance. A similar idea was recently investigated by Ref.~\cite{Man:2025zfu}, which finds that the mixing between  $\psi(4S)$ and $\psi(3D)$ via their couplings to the open-charm channels can bring down their masses to be much lower values than the bare masses of the potential quark model prediction. Two eigenvalues at $4236$ and $ 4387$ MeV are extracted which can be assigned to the experimental observations in the corrsponding mass region, i.e. $\psi(4230)$ and $\psi(4360)$. However, in such a mixing picture one has to explain why the production of $Z_c(3900)$  is favored in the decay of $\psi(4230)$ as a superposition of $\psi(4S)$ and $\psi(3D)$.  

In order to have a better understanding of different theoretical solutions, we focus on several popular phenomenological approaches to highlight their dynamical aspects as follows.

\subsection{The compact tetraquark picture}
The compact tetraquark picture is an extension of normal mesons and baryons within the framework of the quark model. 
For a compact multiquark state, there are various configurations for the compact multiquark wave functions. 
For simplicity, the diquark-antidiquark picture is usually adopted for describing the observed tetraquark candidates. It was proposed that $\psi(4230)$ should be the first orbital excitation compact tetraquark $(cs)(\bar{c}\bar{s})$ based on the diquark-antidiquark picture~\cite{Maiani:2005pe} shortly after its observation. 

\begin{figure}
    \centering
\includegraphics[width=0.45\linewidth]{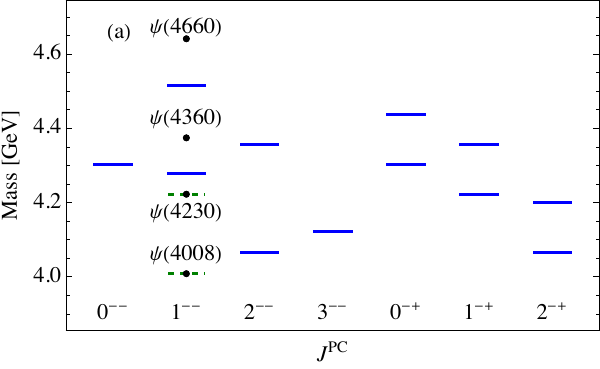}\includegraphics[width=0.45\linewidth]{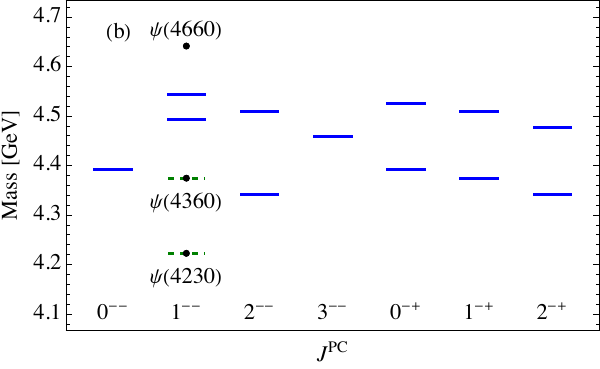}
    \caption{The updated $P$-wave tetraquark spectroscopy (i.e. Fig.2) in Ref.~\cite{Cleven:2015era}. Green dashed and blue solid lines are for inputs and predictions, respectively. Figure (a) is the spectroscopy with $\psi(4008)$ and $\psi(4230)$ as inputs. 
    Figure (b) is the spectroscopy with $\psi(4230)$ and $\psi(4360)$ as inputs. }   
    \label{fig:tetraquark}
\end{figure}

\begin{figure}
    \centering
\includegraphics[width=0.45\linewidth]{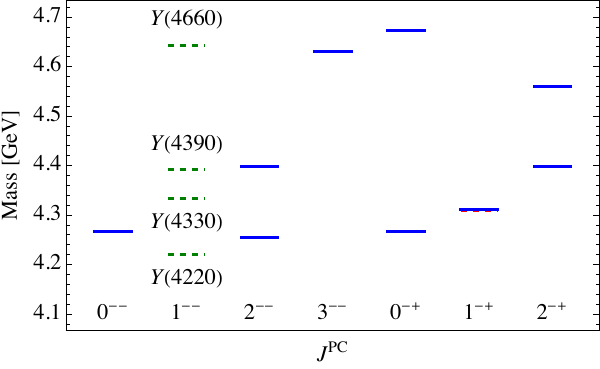}
    \caption{The $P$-wave diquark-antidiquark tetraquark supermultiplet with tensor force. The figure is plotted based on Tab.~VI of Ref.~\cite{Ali:2017wsf}. The green dashed and blue solid curves are the inputs and predictions, respectively. The two $1^{-+}$ states are almost degenerate to each other, with one of them illustrated by the red dashed line. }   
    \label{fig:tetraquark_tensor}
\end{figure}

In the diquark-antidiquark picture a tetraquark mass is given by~\cite{Maiani:2014aja} 
\begin{eqnarray}
    M=M_{00}+B_c \frac{\boldsymbol{L}^2}{2}-2 a \boldsymbol{L} \cdot \boldsymbol{S}+2 \kappa_{c q}\left[\left(\boldsymbol{s}_q \cdot \boldsymbol{s}_c\right)+\left(\boldsymbol{s}_{\bar{q}} \cdot \boldsymbol{s}_{\bar{c}}\right)\right]
\end{eqnarray}\label{mass-rela-tetraquark}
with the parameters $M_{00}, B_c, a$ and $\kappa_{c q}$ fixed from experiment. A compact tetraquark can be expressed by $|s, \bar{s} ; S, L\rangle_J$ with $s$ ($\bar{s}$) the spin of the diquark (antidiquark) and their total spin $S$. 
$L$ is the relative orbital angular momentum between the diquark and antidiquark. 
Note that the total spin of $c\bar{c}$ in this Hamiltonian is not fixed. As a result, the coupling strength between the vector $\psi(4230)$ and $e^+e^-$ is also not fixed. 
In total, there will be four $1^{--}$ and two $1^{-+}$ $P$-wave tetraquarks as partners~\cite{Maiani:2014aja,Cleven:2015era}~\footnote{A QCD sum rule analysis~\cite{Wang:2018ejf} also obtain the similar mass spectrum based on the diquark-antidiquark picture.}, as shown in Fig.~\ref{fig:tetraquark}~\footnote{The spectrum of the mixture picture~\cite{Dias:2012ek} between the compact tetraquark and normal charmonium is not discussed here. Ref.~\cite{Albuquerque:2015nwa} calculates the branching ratio of $\mathcal{B}(B\to \psi(4230)K)=(1.34\pm 0.47)\times 10^{-6}$ based on the mixture of charmonium and exotic tetraquark $[cq][\bar{c}\bar{q}]$.}. An updated mass formula is presented in Ref.~\cite{Ali:2017wsf}, where the tensor force is included. The corresponding mass spectrum is illustrated in Fig.~\ref{fig:tetraquark_tensor}.  

Figure~\ref{fig:tetraquark} is the updated spectroscopy of the $P$-wave tetraquarks based on the mass relation of Eq.~(\ref{mass-rela-tetraquark}) from Ref.~\cite{Maiani:2005pe}. 
The newly averaged mass values of $\psi(4230)$ and $\psi(4360)$ are used in the calculation. In Ref.~\cite{Maiani:2005pe} $\psi(4008)$ is assigned as the lowest tetraquark state of four associated $1^{--}$ states. 
However, the recent BESIII measurement~\cite{BESIII:2022qal} seems to have excluded 
the existence of $\psi(4008)$. For comparison, we consider both cases, namely,
$\psi(4008)$ either exists or not, to investigate the $P$-wave tetraquark spectrum. 
Figure~\ref{fig:tetraquark}(a)
indicates that with $\psi(4008)$ and $\psi(4230)$ as inputs, the mass positions of $\psi(4360)$ and $\psi(4660)$ cannot be accommodated into the $P$-wave tetraquarks. In the second case, supposing $\psi(4008)$ does not exist, one can then use $\psi(4230)$ and $\psi(4360)$ as inputs (Fig.~\ref{fig:tetraquark}(b)). However, $\psi(4660)$ still turns out to be far away from the predicted pattern. In both cases, the mass splittings between the two states for quantum numbers $0^{-+}$, $1^{-+}$, $2^{-+}$ are equal to each other. Once the tensor force is included~\cite{Ali:2017wsf} as shown in Fig.~\ref{fig:tetraquark_tensor}, the mass splittings for those three quantum numbers are not equal any more. Specifically, the two $1^{-+}$ states almost coincide with each other, while the mass difference between two energy levels for $0^{-+}$ are significantly enlarged due to the tensor force~\cite{Ali:2017jda}. 

Based on the tetraquark scenario, it is interesting to note that tetraquarks of  the exotic quantum numbers of $J^{PC}=1^{-+}$ can be accessed. Two states around $4.3~\gev$ and $4.4~\gev$, respectively, are predicted. As shown in  Fig.~\ref{fig:tetraquark}, one can see that
the mass spectrum in the tetraquark picture is much richer than those in the hadro-quarkonium and the $D_1\bar{D}$ hadronic molecular pictures~\cite{Cleven:2015era}. In addition, the isovector partners also  exist for all isoscalar states~\cite{Cleven:2015era}, which makes the spectrum more complicated than that measured in experiment.  
A recent quark model calculation seems to favor  $\psi(4230)$ as a $P$-wave tetraquark ~\cite{Zhao:2025kno}. 

The compact tetraquark $\psi(4230)$ is also expected to be produced in the non-leptonic $B$ decays~\cite{Bigi:2005fr}, which should be sensitive to the quark contents of $\psi(4230)$. Depending on whether $\psi(4230)$ is a conventional charmonium, a hadronic molecule, or a compact tetraquark, its production rate in $B_s\to \psi(4230)\phi$ will differ significantly.  Theoretical studies of the production rates based on these different scenarios may provide criteria to distinguish them with the help of experimental measurements. For instance, for an isoscalar tetraquark of $c\bar{c}(u\bar{u}+d\bar{d})/\sqrt{2}$ or $c\bar{c}s\bar{s}$, its production can be through the $c\bar{c}$ component production via $b\to c (\bar{c} s)$ which is singly Cabibbo suppressed, or via $b\to c (\bar{c} d)$ which is doubly Cabibbo suppressed. For an isovector $\psi(4230)$, its production in the $B/B_s$ decays will require to combine the light quark contents. Namely, its production with an additional light quark-antiquark pair creation should be necessary.

\begin{figure}
    \centering
\includegraphics[width=0.6\linewidth,angle=270]{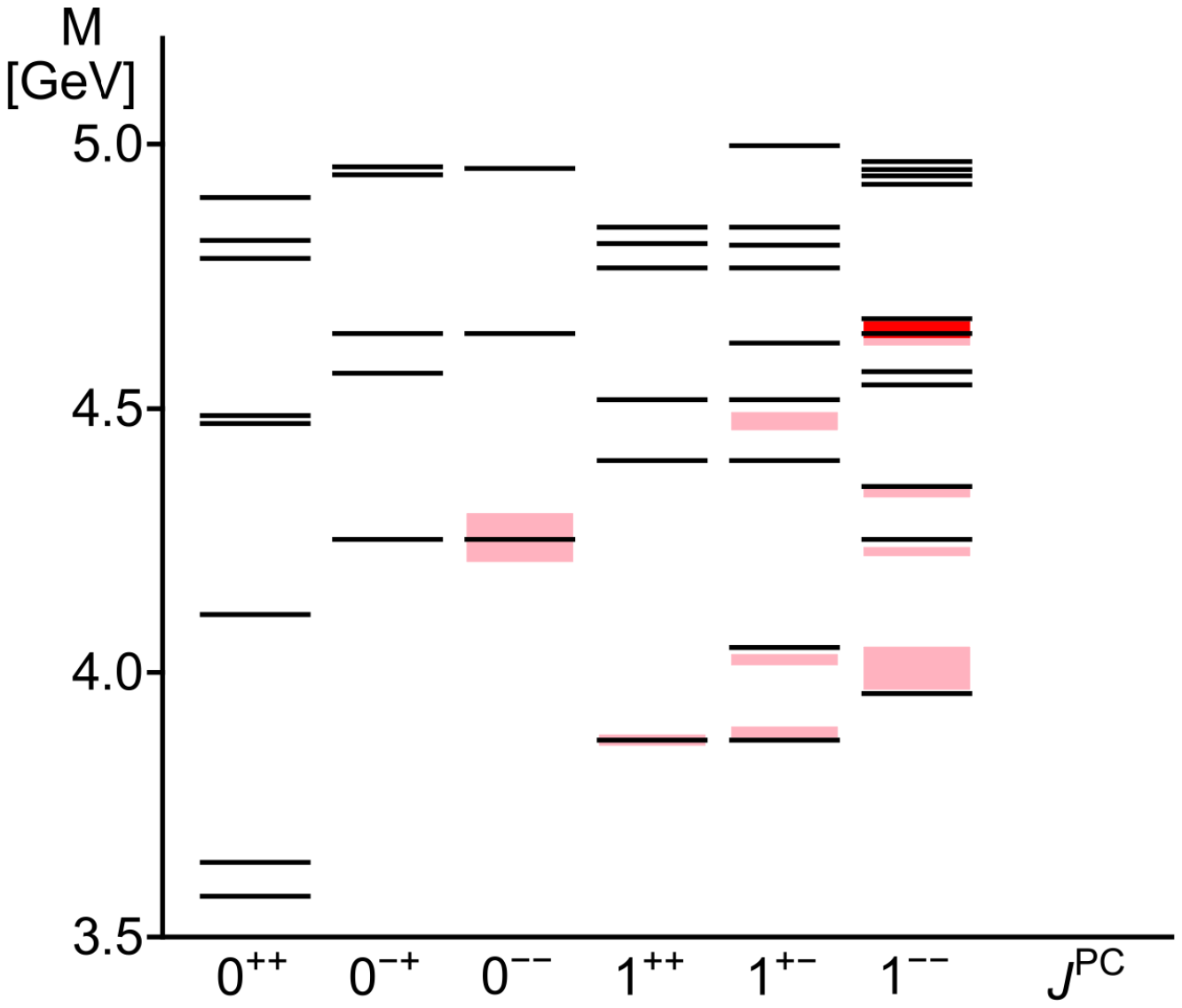}
    \caption{The figure is from  Ref.~\cite{Anwar:2018sol}, which present the $cq\bar{c}\bar{q}$ tetraquark spectrum (black solid lines in comparison with the existing experimental candidates (red box)). The linear confinement potential and one-gluon-exchange potential are considered in this calculation.}   
    \label{fig:tetraquark2}
\end{figure}

At present, studies of compact multiquark states are more focused on the mass spectrum. 
Theoretical calculations of the tetraquark spectroscopy based on various potentials can be found in the literature. For instance, Ref.~\cite{Anwar:2018sol} studies the spectroscopy of tetraquarks with quark content $c\bar{c}q\bar{q}$ based on the one-gluon-exchange potential plus the linear confinement potential.  
The predicted spectroscopy is presented in Fig.~\ref{fig:tetraquark2}. The lower few states have a similar energy pattern as that on the left panel of  Fig.~\ref{fig:tetraquark}. 
One notices that the mass splitting among higher states becomes smaller. Taking into account the broadness of higher states, such a behavior will bring big challenges for experimental detection. 

It should be noted that in recent years various numerical methods have been developed for quantitative study of multiquark states. In particular,  the Gaussian expansion method (GEM)~\cite{Hiyama:2003cu}, which was developed in nuclear physics, is applied to the study of hadron spectroscopy with heavy flavors and no diquark assumption~\cite{Liu:2019zuc,liu:2020eha,Liu:2021rtn,Deng:2023mza,Meng:2023for,Yang:2023mov}. In Ref.~\cite{Wang:2022yes}, a complex scaling method is constructed which turns out to be powerful for precisely locating resonances on the complex energy plane.

\subsection{The hybrid picture}
Hybrid meson is referred to a color neutral object within which the quark and antiquark system is in the irreducible color-octet representation, and the gluonic degrees of freedom will be  explicitly combined together with the quark-antiquark to form a color singlet. In the flux tube picture the gluon motion in a hybrid can have two distinct modes~\cite{Isgur:1983wj,Isgur:1984bm}. One is the transverse mode where the gluon behaves like a constituent gluon. The other is the collinear mode where the gluon motion is along the direction between the constituent quarks and provides an effective potential similar to that for quark and antiquark in a conventional meson.
These two different modes actually give different mass splittings between the ground states and the first excited states of charmonium hybrids. The transverse mode can cause a mass difference for about  $1.2\sim 1.4~\gev$ between the ground and first excited states as shown by the LQCD simulation~\cite{Ma:2019hsm}. In contrast, the collinear mode
will cause much smaller mass differences.  

The assignment of $\psi(4230)$ as a hybrid was proposed by Refs.~\cite{Zhu:2005hp,Close:2005iz,Kou:2005gt} at the early stage. In the hybrid picture its coupling to a di-lepton pair is highly suppressed~\cite{Close:2005iz} due to its additional gluon component. It is expected to be smaller than the di-lepton coupling of the $S$-wave charmonia, but larger than that of the $D$-wave ones. This may explain  its small leptonic decay width, $\Gamma_{ee}(\psi(4230))$, and possible small cross sections around $4.23~\gev$ in $e^+e^-\to \mathrm{hadrons}$. Meanwhile, a selection rule in the picture of the flux tube model suggests that the ground state vector charmonium hybrid decays into a pair of ground-state $S$-wave charmed mesons of the same spatial size will be suppressed~\cite{Close:2005iz,Isgur:1983wj,Isgur:1984bm}. This selection rule manifests itself for heavy quarkonium hybrids in the heavy quark limit. However, due to the heavy quark symmetry (HQS) breaking for the charmonium system, e.g. $m_{D^*}-m_D\simeq 152$ MeV, this selection rule actually does not hold strictly. Nevertheless, if threshold effects become important, this selection rule will also break down~\cite{Close:2005iz}. 

A recent analysis based on the Born-Oppenheimer (BO) approximation indicates a selection rule different from the flux tube model. By assigning $\psi(4230)$ as the lowest vector charmonium hybrid, its relative partial decay rates to the $S$-wave open charmed pair is predicted to be $\Gamma(\psi(4230)\to D\bar{D}):\Gamma(\psi(4230)\to D^*\bar{D}+c.c.):\Gamma(\psi(4230)\to D^*\bar{D}^*)=1:0:3$~\cite{Braaten:2024stn}, where only $\psi(4230)\to D^*\bar{D}+c.c.$ is forbidden. In contrast, its partners, i.e. $(0,1,2)^{-+}$ will equally decay to $D^*\bar{D}+c.c.$ and $D^*\bar{D}^*$. Note that states of $(0,2)^{-+}$ cannot decay into two pseudoscalars due to spin-parity conservation. For $1^{-+}$ states they cannot decay into two identical pseudoscalars because of Bose statistics.    

The selection rule with the  BO approximation for the couplings of heavy charmonium hybrids to the $S$-wave charmed meson pairs is extracted in the HQS limit~\cite{Bruschini:2023tmm,Shi:2023sdy}. In reality the selection rule may also break down for hybrids with conventional quantum numbers. Given that soft-gluon exchanges between the heavy quark-antiquark pair and the light quark-antiquark pair created from the gluon lump are inevitable, as a result, it would be difficult to tell the dynamic difference between a hybrid decaying into open channels with a quark-antiquark pair created through the constituent gluon and a conventional meson decaying into open channels via a quark pair creation model (i.e. $^3P_0$ model). 
It should also be noted, for heavy quarkonium hybrids, the implementation of the BO approximation leads to much smaller mass differences between the ground states and the first excited ones~\cite{Braaten:2014qka}. 

\begin{figure}
    \centering
\includegraphics[width=0.5\linewidth]{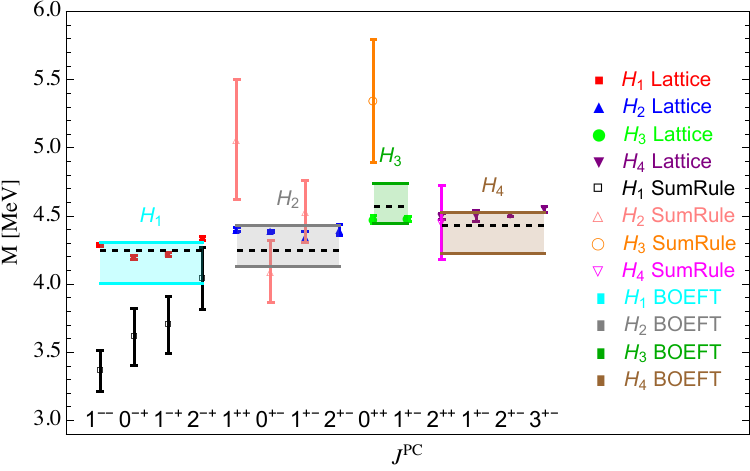}
    \caption{The figure is a combination of Fig.4, Fig.5 and Fig.7 of Ref.~\cite{Berwein:2015vca}. 
    The red box, blue triangle, green circle and purple inverted triangle are the Lattice calculations~\cite{HadronSpectrum:2012gic} for the $H_1$, $H_2$, $H_3$, $H_4$ hybrids, respectively. The naming scheme $H_i$ follows that in Refs.~\cite{Juge:1999ie,Braaten:2014qka,Braaten:2014ita,Braaten:2013boa}, where $H_1$ ($H_2\cup H_3\cup H_4$) is the hybrid with a combination of the $J^{PC}=1^{+-}$ glue-lump and a pair of $S$-wave ($P$-wave) heavy quark and antiquark. The black hollow box, pink hollow triangle, orange hollow circle and Magenta inverted triangle are the QCD sum rule results~\cite{Chen:2013zia}. The black dashed lines are the charmonium hybrid multiplets from Refs.~\cite{Braaten:2014qka,Braaten:2014ita,Braaten:2013boa} without the hyperfine splitting among the states within a given multiplet. The light blue, gray, darker green and brown bands are the results from nonrelativistic effective field theory based on the BO approximation~\cite{Berwein:2015vca}, 
  where the static heavy quark pair serves as a color source and the gluon adjusts its configurations to fit a given quantum number.}   
    \label{fig:hybrid}
\end{figure}

A vector charmonium hybrid ground state actually implies the existence of its quark-spin-triplet partners.
The calculation of hybrids in lattice QCD was inspired by the flux tube model~\cite{Isgur:1984bm}. An unquenched lattice QCD calculation~\cite{HadronSpectrum:2012gic} concludes that the $1^{-+}$ charmonium-like hybrid should dominantly decay into open charm modes, such as $D_1\bar{D}$, $D^*\bar{D}$, $D^*\bar{D}^*$. 
The mass position of $\psi(4230)$ agrees well with one of the $1^{--}$ hybrid~\cite{HadronSpectrum:2012gic}. Also, it shows that the lightest $1^{-+}$ hybrid is about $100~\mev$ below $\psi(4230)$~\cite{HadronSpectrum:2012gic} (see Fig.~\ref{fig:hybrid}). 

Recently, updated results of the BO hybrids are presented in Ref.~\cite{Berwein:2024ztx}. However, it does not include the whole hybrid spectrum. In Fig.~\ref{fig:hybrid} we present the hybrid spectrum of its first version~\cite{Berwein:2015vca} as a comparison. A quenched lattice NRQCD calculation qualitatively predicts that the mass ordering of these partner states reads $0^{-+}<1^{-+}<1^{--}<2^{-+}$ for the $b\bar{b}g$ hybrid~\cite{Drummond:1999db}, which is in agreement with that in Ref.~\cite{Kalashnikova:2008qr}. The flux-tube model 
predicts that the coupling of the $D$-wave $1^{--}$ charmonium hybrid to one $P$-wave $s_l=\frac{3}{2}^+$ and one $S$-wave $s_l=\frac{1}{2}^-$ channel is small~\cite{Close:1994hc,Page:1998gz}. This feature is significantly different from that of the molecular picture, which should be useful for distinguishing these two scenarios.  

\subsection{The hadro-quarkonium picture}
\begin{figure}
\centering
\includegraphics[width=0.45\linewidth]{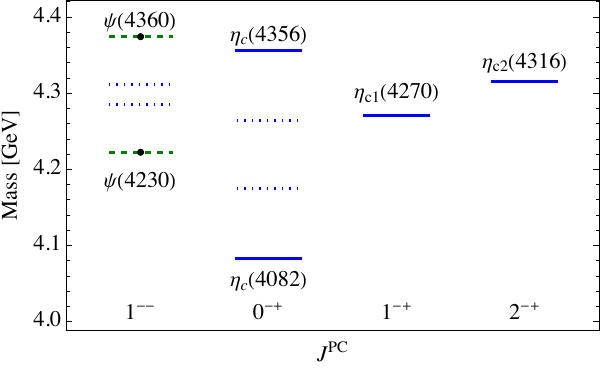}
\caption{The updated hadro-quarkonium spectroscopy, i.e. Fig.2 in Ref.~\cite{Cleven:2015era}. The green dashed lines are the experimental inputs. The two blue dotted lines with quantum number $1^{--}$ (or $0^{-+}$) represent the masses of the unmixed states. The blue solid lines are the predicted  partners. }   
\label{fig:hadroquarkonium}
\end{figure}

Based on the experimental fact that most of vector charmonium-like states are observed in the final states with one charmonium plus several light hadrons, Voloshin proposed the ``so-called" hadro-quarkonium picture~\cite{Voloshin:2007dx}. It is a system with compact heavy quarkonium embedded inside a light quark cloud~\cite{Voloshin:2007dx,Dubynskiy:2008mq}. For instance, $\psi(4230)$ can be considered as one of the $1^{--}$ hadro-charmonia, with the mixture to another hadro-charmonium $\psi(4360)$ observed in the $h_c\pi\pi$ channel~\cite{Li:2013ssa,Cleven:2015era}. They are considered as mixtures of two hadro-charmonia $(1^{--})_{c\bar{c}}\otimes (0^{++})_{q\bar{q}}$ and $(1^{+-})_{c\bar{c}}\otimes (0^{-+})_{q\bar{q}}$ with their decays into $J/\psi\pi\pi$ and $h_c\pi\pi$, respectively, due to HQSS and HQS. In the hadro-charmonium picture, with $\psi(4230)$ and $\psi(4360)$ as input, the predicted $1^{-+}$ partner is located around $4.27~\gev$ as shown by Fig.~\ref{fig:hadroquarkonium}. To some extent, the hadro-quarkonium picture has a similar description of the gluon and light quark degrees of freedom as the BO model while the heavy quark-antiquark core plays a role as the static color source. 

A lattice QCD simulation of the binding energies between quarkonium and the light hadrons suggests that the binding energy is similar to the deuterium binding energy, thus, will make it a loosely bound state~\cite{Alberti:2016dru}. A calculation based on the chromo-electric polarizability in the framework of the $1/N_c$ expansion  is performed in Ref.~\cite{Ferretti:2018kzy}, 
where $\psi(4230)$ is also proposed to be a vector hadro-quarkonium. Ref.~\cite{MartinContreras:2023oqs} uses the softwall model in AdS/QCD to explain the exotic spectroscopy in various scenarios. They also conclude that $\psi(4230)$ is more acceptable in the hadro-quarkonium scenario.  

In Ref.~\cite{MartinezTorres:2009xb}, the $\psi(4230)$ was dealt within the Faddeev equations with the coupled-channels $J/\psi K\bar{K}$, $J/\psi\pi\pi$, $D^*\bar{D}\pi$ and others. In addition, the $D^*\bar{D}$, $J/\psi\pi$ and other coupled channels were considered in Ref.~\cite{Aceti:2014uea} to explain the $Z_c(3900)$, and the mass distribution for the $e^+e^-\to D^*\bar{D}\pi+c.c.$ process could be well reproduced. The combination of these two works shows that the picture of Ref.~\cite{MartinezTorres:2009xb} can lead to the production of $Z_c(3900)$ in the $\psi(4230)\to Z_c(3900)\pi\to J/\psi\pi\pi$ process, and also account for the $\psi(4230)\to J/\psi f_0(980)$ decay (with the $f_0(980)$ made up from the $\pi\pi$ and $K\bar{K}$ channels), another of the reported decay modes of the $\psi(4230)$ resonance. However, the partial decay rates to these channels were not evaluated in these works.


\subsection{The hadronic molecule picture}

Shortly after the observation of $\psi(4230)$ (i.e. $Y(4260)$) in 2005, there were proposals for explaining it as hadronic molecules, such as $\rho \chi_{c1}$~\cite{Chiu:2005ey,Liu:2005ay}, $\omega\chi_{c1}$~\cite{Yuan:2005dr,Dai:2012pb}, $\Lambda_c\bar{\Lambda}_c$~\cite{Qiao:2005av}. Later, it was pointed out in Ref.~\cite{Swanson:2005tq} that $\psi(4230)$ could be an isospin-singlet hadronic molecule made of $D_1\bar{D}+c.c.$ since it is close to the first $S$-wave threshold $D_1\bar{D}$ in the vector charmonium spectrum.
In Ref.~\cite{Ding:2008gr} the scenario of $D_1\bar{D}+c.c.$ hadronic molecule was first quantitatively investigated with a coupled-channel formalism for the $D_1\bar{D}+c.c.$ and $D_0\bar{D}^*+c.c.$ interactions except that both $D_1$ and $D_0$ are broad~\footnote{In Refs.~\cite{Close:2010wq,Close:2009ag}  $\psi(4230)$ is described as a deeply bound state formed by the $\bar{D}$ meson and the broader $D_1$ (i.e. $D_1(2430)$) with the light degree of freedom $s_l=\frac{1}{2}^+$ through the one-pion-exchanged (OPE) potential. The large width of $D_1^\prime\to D^*\pi$ gives a strong coupling $g_{D_1^\prime D^*\pi}$, which will deduce a strong OPE potential. However, in this situation a proper treatment of three-body unitarity~\cite{Filin:2010se} will dismiss the significance of the peak structure. It suggests that a coupled-channel analysis including $D_1(2420)\bar{D}+c.c.$, $D_1(2430)\bar{D}+c.c.$ and $D_0\bar{D}^*+c.c.$ should be necessary.}. 

After the observation of the charged hidden charm tetraquark state $Z_c(3900)$~\cite{BESIII:2013ris,Belle:2013yex,Xiao:2013iha} it was proposed by Ref.~\cite{Wang:2013cya} that $\psi(4230)$ could be a hadronic molecule made of $D_1(2420)\bar{D}$, where $D_1(2420)$ is the narrow $D_1$ with the HQSS configuration $|1^+,j_l^p=\frac{3}{2}^+\rangle$. 
In Refs.~\cite{Wang:2013kra,Cleven:2013mka} the molecular picture was investigated in detail and a pole mass of $4217.2\pm 2.0$ MeV was extracted for the first time by studying the cross section lineshapes of $e^+e^-\to J/\psi\pi\pi$, $h_c\pi\pi$ and $D\bar{D}^*\pi$. 

Recently, global analyses of a series of vector charmonium-like states are carried out based on  the high-statistic BESIII measurement of the $e^+e^-$ annihilation cross sections~\cite{vonDetten:2024eie,Nakamura:2023obk}. The extracted pole positions are consistent with each other, i.e.
$4227\pm 4-i (25^{+4}_{-1})~\mathrm{MeV}$ from Ref.~\cite{vonDetten:2024eie} and $4229.9\pm 0.9-i (23.2\pm 1.3)~\mev$ from Ref.~\cite{Nakamura:2023obk}. Both poles are about $60~\mev$ below the $D_1\bar{D}$ threshold, making $\psi(4230)$ a good candidate as an isospin-singlet $D_1\bar{D}$ hadronic molecule.

\begin{figure}
    \centering
\includegraphics[width=0.5\linewidth]{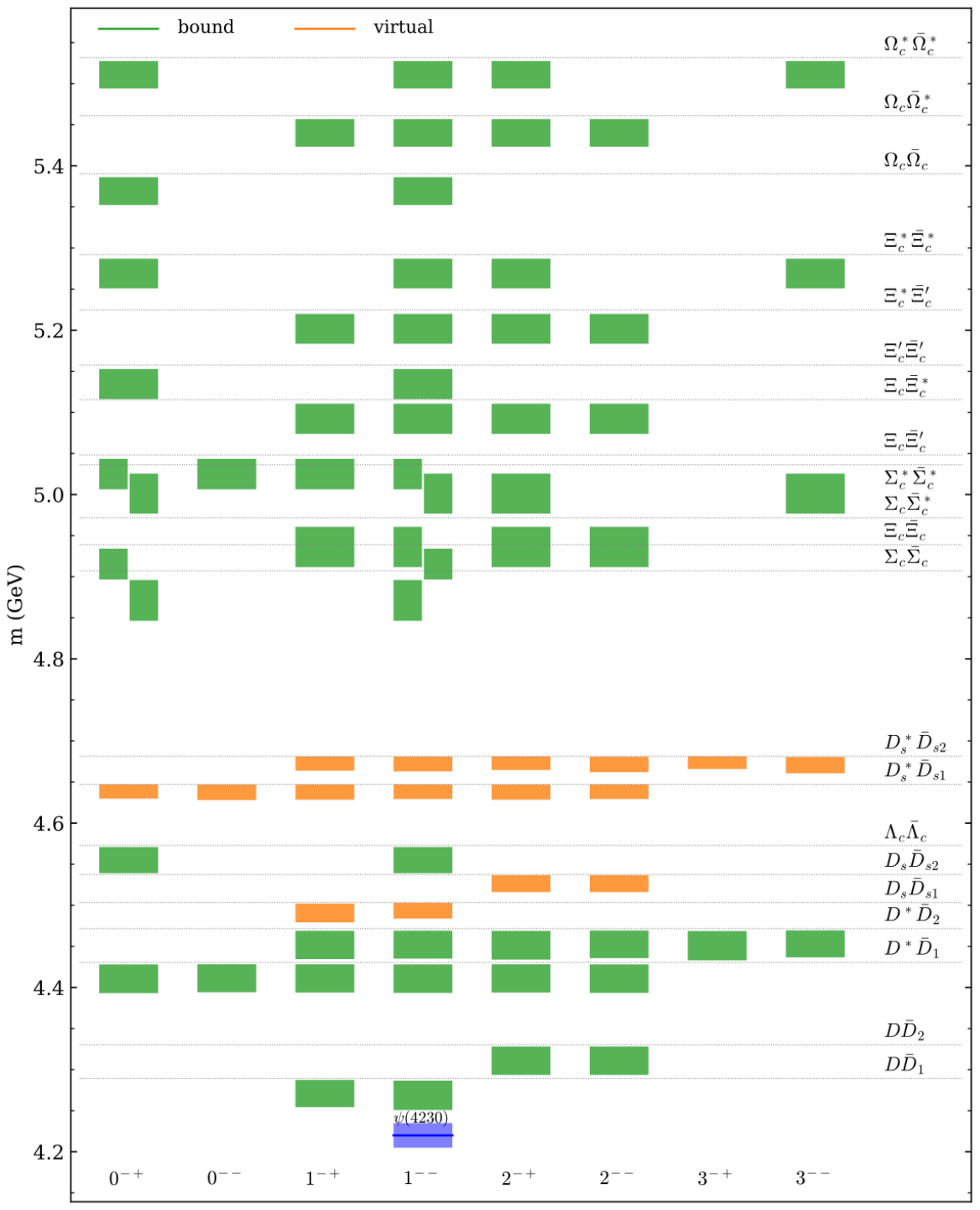}
    \caption{Figure taken from Ref.~\cite{Dong:2021juy}, which shows the spectrum of the isospin singlet hidden charm hadronic molecules with strangeness zero. The green and orange bands are for bound and virtual states, respectively. The blue line (band) represents the central value (error) of $\psi(4230)$ from the PDG~\cite{ParticleDataGroup:2020ssz}. }   
    \label{fig:HM}
\end{figure}

During the past decade, intensive studies of $\psi(4230)$ based on the molecular picture were carried out to explore various possible consequences~\cite{Wang:2013cya,Li:2013bca,Liu:2013vfa,Li:2013yla,Wang:2013kra,Cleven:2013mka,Dong:2013kta,Wu:2013onz,Li:2014gxa,Qin:2016spb,Chen:2016byt,Wang:2016wwe,Cleven:2016qbn,Lu:2017yhl,Xue:2017xpu,Chen:2019mgp,Peng:2022nrj,Qin:2016spb,Li:2014gxa,Guo:2014ura,Peng:2021hkr}. The $D_1\bar{D}$ molecular picture can well describe its lineshapes in $J/\psi\pi\pi$, $h_c\pi\pi$ and $D^0D^{*-}\pi^+$ channels ~\cite{Cleven:2013mka}. The further combined analysis~\cite{Qin:2016spb,Xue:2017xpu} also confirms its molecular scenario. A dispersive analysis of both the $\psi(4230)\to J/\psi\pi^+\pi^-$ and $\psi(4230)\to J/\psi K^+K^-$ processes indicates its sizable non-molecular component from the light quark perspective~\cite{Chen:2019mgp}.
As follows, we list the main consequences of the hadronic molecule scenario which may serve as evidence for its hadronic molecule nature:

\begin{itemize}


\item Being the $D_1(2420)\bar{D}$ molecule with $D_1(2420)$ of the HQSS configuration $|1^+,j_l^p=\frac{3}{2}^+\rangle$, the virtual photon coupling of the $D_1\bar{D}$ pair is suppressed by factor 
$\frac{(E-2m_c)^2}{2(E+m_c)^2}$ 
with $E$ the center-of-mass energy~\cite{Li:2013yka,Wang:2013kra,Qin:2016spb}. It implies a significantly small di-lepton decay width of $\psi(4230)$. Furthermore, as the $D_1(2420)\bar{D}$ molecule, its dominant decay channel will be $D\bar{D}^*\pi+c.c.$~\cite{Cleven:2013mka,vonDetten:2024eie,Nakamura:2023obk,Qin:2016spb}, while its two-body open charm decays will be suppressed.
These features explain the insignificant signal of 
$\psi(4230)$ in the two-body open channels and $e^+e^-\to \mathrm{hadrons}$~\cite{Wang:2013kra}.  

\item Based on the molecular picture the largely asymmetric cross section lineshape observed in $e^+e^-\to J/\psi\pi\pi$ can be naturally explained by only one state instead of two~\cite{Cleven:2013mka,vonDetten:2024eie,Nakamura:2023obk,Xue:2017xpu,Cleven:2016qbn,Qin:2016spb,Wu:2013onz}. Meanwhile, it shows that $Z_c(3900)$ can also be explained as the $D\bar{D}^*$ hadronic molecule state, which is enhanced by the triangle singularity mechanism in the $\psi(4230)$ decays~\cite{Wang:2013cya,Wang:2013hga,Liu:2015taa}. Such a correlation also predicted an enhanced production of $\chi_{c1}(3872)$ (i.e. $X(3872)$) in $e^+e^-\to \psi(4230)\to \gamma \chi_{c1}(3872)$~\cite{Guo:2013zbw} which was confirmed by the BESIII measurement~\cite{BESIII:2020nbj}.

\item Since the charge conjugate transformation of the $D_1\bar{D}$ pair is not itself, but its anti-particle pair $\bar{D}_1D$, their linear combination can give both $J^{PC}=1^{--}$ and $1^{-+}$ states with wave functions $\frac{1}{\sqrt{2}}\left(D_1\bar{D}- \bar{D}_1D\right)$ and $\frac{1}{\sqrt{2}}\left(D_1\bar{D}+ \bar{D}_1D\right)$, respectively. Their dynamics can be described by the same set of parameters based on the HQSS. In this sense, one can predict the existence of the $1^{-+}$ partner of $\psi(4230)$~\cite{Wang:2014wga,Dong:2019ofp,Zhang:2025gmm}. Based on the light vector-meson-exchanged model the exotic $1^{-+}$ states and  other hidden charm exotics are investigated by Ref.~\cite{Dong:2021juy} (see Fig.~\ref{fig:HM}). It shows that one of the best channels to look for these exotic candidates would be via the radiative decay of the corresponding vector charmonium-like states~\cite{Wang:2014wga,Dong:2019ofp,Zhang:2025gmm}. This study can also be extended to scattering between the two doublets---$(D_1, D_2)$ and $(\bar{D}, \bar{D}^*)$---revealing additional partner states. Among these, a robustly predicted flavor-neutral charmonium-like exotic state with $J^{PC}=0^{--}$, designated $\psi_0(4360)$ based on its mass, emerges~\cite{Ji:2022blw}. Its best-measured production channel, $e^+e^- \to \eta\psi_0(4360)$, is uniquely identified through the angular distribution of the outgoing $\eta$ meson~\cite{Ji:2022blw}.


\end{itemize}

\section{Summary and Outlook}

In this short review we present the status of experimental and theoretical progress in the study of $\psi(4230)$. We stress that the direct production of $\psi(4230)$ in $e^+e^-$ annihilation can provide more observables for understanding the nature of $\psi(4230)$, which include the cross section lineshape,  decay widths to exclusive channels, di-leptonic decay width, etc. We also review the possible theoretical interpretations in the literature. Apart from a general survey, we focus on four scenarios, i.e. compact tetraquark, charmonium hybrid, hadro-charmonium, and hadronic molecule, regarding their broad coverages in hadron physics. It shows that these models have tackled some features of the $\psi(4230)$ from different view angles, and are also driven by the experimental observations.

One crucial issue arising from all these possible solutions is that there will be associated consequences from each solution. Instead of a single exotic state, one may expect a complete spectrum or associated partner states which are driven by the same dynamics. Note the fact that although there have been tens of exotic candidates observed in experiment, the number of states is still far less than what we expect from a multiquark system. In this sense, the hadronic molecule picture provides an economic solution for most of the signals observed in the vicinity of $S$-wave thresholds. For more complicated multiquark spectra one still needs to understand ``How some of those multiquark systems can be stabilized, but not the others?", or ``To what extent, we can expand the multiquark spectra and where to find them?"

So far, the signals of $\psi(4230)$ are mainly from $e^+e^-$ annihilation, and its signals in other processes, such as the $B$ decays, still need confirmation. In heavy hadron weak decays we would expect that different production mechanisms and different consequences arise from different scenarios. This makes the weak processes another interesting probe for studying the nature of $\psi(4230)$. For instance, one possible process to search for $\psi(4230)$ is via $B_s\to \psi(4230)\phi$, where $\psi(4230)$ can be produced via its $c\bar{c}$ component.
Depending on whether the $\psi(4230)$ is a conventional charmoinium, a hadronic molecule, or a compact tetraquark, its production rate in the $B_s\to \psi(4230)\phi$ will differ significantly. Theoretical studies of the production rates based on these different scenarios may provide criteria to distinguish them with the help of experimental measurements. 
 Note that both $\psi(4230)$ and $\phi$ can be searched for in their di-leptonic decays into $2(\mu^+\mu^-)$ at CMS or in the final state of $\mu^+\mu^- K^+ K^-$ at LHCb.  If it is a hadronic molecule state, perhaps both the intermediate $c\bar{c}$ and $D_1\bar{D}$ production can become crucial. For instance, the $\psi(4230)$ production can occur via triangle diagrams~\cite{Liu:2024hba,Yuan:2025pnt}. In brief, future high-statistics data from experiment are still needed in order to allow us to gain deeper insights into the nature of $\psi(4230)$.

\vskip 2mm
We are grateful for the inspiring and beneficial collaborative works with Vadim Baru, Martin Cleven, Leon v. Detten, Meng-Lin Du, Feng-Kun Guo, Christoph Hanhart, Gang Li, Xiao-Hai Liu, Ulf-G. Mei{\ss}ner, Alexey Nefediev, Wei Wang, and Bing-Song Zou, on the relevant topics. 
We thank Ying Chen and Liuming Liu for useful discussions on the lattice QCD simulations. We also thank Yu-Ping Guo, Hai-Bo Li, Rong-Gang Ping, Cheng-Ping Shen, Xiao-Yan Shen, and Chang-Zheng Yuan for many useful discussions on the experimental results. Special ackowledgements to Ying Chen, Feng-Kun Guo, Christoph Hanhart, and Chang-Zheng Yuan for their careful readings and feedbacks on the manuscript.
This work is supported in part
by the NSFC under Grants Nos.~12375073 and 12235018.


\end{document}